*Article*

# Design and Implementation of a Novel Compatible Encoding Scheme in the Time Domain for Image Sensor Communication

**Trang Nguyen, Mohammad Arif Hossain and Yeong Min Jang \***

Department of Electronics Engineering, Kookmin University, Seoul, 02707, Korea; trang@kookmin.ac.kr (T.N.); dihan.kuet@gmail.com (M.A.H.)
**\*** Correspondence: yjang@kookmin.ac.kr (Y.M.J); Tel.: +82-2-910-5068.

**Abstract:** This paper presents a modulation scheme in the time domain based on On-Off-Keying and proposes various compatible supports for different types of image sensors. The content of this article is a sub-proposal to the IEEE 802.15.7r1 Task Group (TG7r1) aimed at Optical Wireless Communication (OWC) using an image sensor as the receiver. The compatibility support is indispensable for Image Sensor Communications (ISC) because the rolling shutter image sensors currently available have different frame rates, shutter speeds, sampling rates, and resolutions. However, focusing on unidirectional communications (i.e., data broadcasting, beacons), an asynchronous communication prototype is also discussed in the paper. Due to the physical limitations associated with typical image sensors (including low and varying frame rates, long exposures, and low shutter speeds), the link speed performance is critically considered. Based on the practical measurement of camera response to modulated light, an operating frequency range is suggested along with the similar system architecture, decoding procedure, and algorithms. A significant feature of our novel data frame structure is that it can support both typical frame rate cameras (in the oversampling mode) as well as very low frame rate cameras (in the error detection mode for a camera whose frame rate is lower than the transmission packet rate). A high frame rate camera, i.e., no less than 20 fps, is supported in an oversampling mode in which a majority voting scheme for decoding data is applied. A low frame rate camera, i.e., when the frame rate drops to less than 20 fps at some certain time, is supported by an error detection mode in which any missing data sub-packet is detected in decoding and later corrected by external code. Numerical results and valuable analysis are also included to indicate the capability of the proposed schemes.

## 1. Introduction

The advent in recent years of optical wireless communication (OWC) has significantly improved communication technology. This mode of communication has led researchers to rethink the history of communication technology. It has been revealed as a blessing for the newly invented Internet of Things (IoT) because to some degree it can help meet the increasing demands for connectivity among things. The scarcity of bandwidth spectrum is a critical issue while connecting thousands of wireless devices in a network, which requires the use of this recent technology to find solutions. The OWC technology is one of the strongest candidates to share the burden of connectivity. The unused optical spectrum is certainly conducive to meeting the demand and can be used without any payment [1].

Consequently, the visible light spectrum is now in a state of action after the IEEE standardization of Visible Light Communication (VLC) technology released in 2011 [2]. Besides, regarding the safety and health of humans, OWC technology has no competitor because it is free from electromagnetic radiation. Moreover, the OWC technology has the advantage of using lighting devices for both illumination and communication purposes, thus enabling its classification as a green communication method. Besides, it is inexpensive and easy to deploy [3,4].

The day is not too far when OWC technology will outshine RF communication technology. Achieving immense success in VLC technology, the most recent standardization activity related to image sensor communication (ISC), a part of OWC technology that uses the camera sensor as a receiver, is the IEEE 802.15.7r1 Task Group (TG7r1), known as the revised VLC specification, which has successfully attracted interest from a large number of researchers from various renowned companies and universities [2], particularly the ISC sub-section. TG7r1 aims to promote the use of OWC technology in commercial products by making it more feasible. The released Technical Consideration Document (TCD) of the TG7r1 mentions the guidelines for the direction of development of any submitted technical proposal [5]. The TCD serves to summarize the applications presented in response to the TG7r1 call for applications as well as the questions and answers. In addition, the principal requirements proposed by the submitted applications have been defined and described by the TCD. It is evident that the revised technological proposals will play a pivotal role in the standardization of the OWC technology. According to the TCD, OWC technology can be classified into three categories: image sensor communications, high-rate photodiode (PD) communications, and low-rate PD communications. ISC uses an image sensor such as a camera as the receiver. The proposed high-rate PD communication is based on bidirectional communication technology, and the receiver utilized in the method is a high-speed PD while low-rate PD communication is a wireless communication technology that uses the PD system in a low-speed photodiode receiver.

The topics of interest of this paper pivot around the ISC technology as the frame rate can vary in the image sensor rather than the PD. Moreover, the compatibility features in supporting various types of rolling shutter cameras are going to be presented. Those features include a support mechanism for solving the problem of the wide frame rate variation observed in plenty of our published works, for instance, a support mechanism for the change in shutter speeds and different sampling rates of different cameras has been discussed in [4], and last but not least a support algorithm for various rolling exposure times and a further transmission distance have also been adopted in this paper. The importance of supporting image sensor compatibility was agreed in the consideration document of TG7r1 as summarized. For compatibility purposes, consideration of the variations in shutter speeds and sampling rates of different rolling shutter cameras is particularly valuable because the available bandwidth is ultra-narrow. A proper bandwidth based on rolling shutter operation which has been found experimentally is recommended in this paper. The compatibility for the dissimilarity in rolling exposure times, defined by the time from the first line of the pixel to the last line of the pixel exposed to light in a rolling shutter image sensor, and the distance can affect the amount of data possible per single image. However, the proposed data frame structures are the fundamental part of the compatibility feature, a novel idea that allows fusing different parts of a sub-packet payload exposed from numerous images into a complete one. The rest of the paper discusses related works in Section 2 while the system analysis has been specified in Section 3. Also, a comparison between our work and the related ones is going to be shown, and then the necessity and the contribution of this work are given as well. The proposed modulation schemes have subsequently been illustrated in the next section. The performance analysis of the proposed system has been entirely explained in Section 5, whereas the conclusions of the work are presented in Section 6.

## 2. Related Works and Our Contributions

### 2.1. Related Works

The most recent activity related to ISC technologies has been published in the proposal presentations of plenty of companies and universities involved in the January 2016 TG7r1

Call-for-Proposals (CfP) meeting. The first CfP meeting witnessed some interesting ISC technology presentations and discussions. Table 1 does not present all of the contributions rather than some of the most relevant technologies discussed. For further details, readers can refer to the submitted presentations available online at [6]. However, a brief review of the contributions on Image Sensor Communications PHY and MAC proposals from Intel, National Taiwan University (NTU), Panasonic, Kookmin University, Seoul National University of Science & Technology (SNUST), and California State University (Sacramento) are given here. Compared to global shutter cameras the popularity of rolling shutter cameras is outstanding. Therefore, taking this into consideration, all of the submitted proposals supported at least a PHY mode for rolling shutter receivers.

The proposals mostly related to rolling shutter modulation schemes include On-Off-Keying (OOK) by Kookmin University, Pulse Width Modulation (PWM) by Panasonic, Multiple-Frequency-Shift-Keying (FSK) by NTU and Kookmin University and Offset-Variable Pulse Position Modulation (VPPM), by SNUST. Meanwhile, a hybrid modulation scheme for either a global shutter camera or a rolling shutter camera or both has been proposed by Intel and Kookmin University in different ways. Specifically, to target high performance in global shutter camera receivers, there is a VPPM modulation scheme suggested by Intel and a dimmable Spatial-PSK modulation scheme proposed by Kookmin University. Last but not least, a high data rate (up to Mbps) MIMO-OCC PHY mode would be continually updated and merged by the concerned proposers.

To compare the aspect of the compatibility support for image sensors, the variation in frame rate camera has been considered with more priority. Our modulation scheme is one of best options for addressing the problem of frame rate variation compared to other proposals and other published works. Roberts proposed an image sensor-based communication system using frequency shift On-Off Keying that can support, however, only a small range of variation in the frame rate [7]. Besides, the scheme can work for fixing or controlling the camera frame rate in a typical image sensor of high quality. Luo et al. have analogously proposed the under-sampled phase shift ON-OFF keying modulation technique including to support non-flickering communication in the VLC technology [8]. However, there is no mention of frame rate variation in the paper. Furthermore, VLC has been proposed by Danakis et al. using Complementary Metal Oxide Semiconductor (CMOS) image sensors, but the authors have not asserted the frame rate variation of the camera sensor [9]. Another research work on VLC using a smartphone camera has been carried out in which the issue of frame rate variation has not been considered either [10]. Langlotz et al. have demonstrated a 2D color code-based ISC method, but does not include the frame rate variation problem [11]. On the other hand, Hu et al. introduced frame rate variation for screen-camera communication [12], but the linearity-ensuring code that has been used in that dissertation is highly complicated due to the simultaneous processing of three images. Besides, the line forward error correction (FEC) code has been proposed to mitigate the frame rate variation issue by Casio. The proposed technique can decrease the data rate and introduce some other limitations [13].

**Table 1.** A survey of modulation schemes.

| Modulation Scheme | Rolling Shutter | Global Shutter |
|---|---|---|
| On-Off-Keying (OOK) | ▪ OOK based RLL coding by PureLiFi [9]<br>▪ Compatible OOK by Kookmin University [6] | ▪ None |
| Frequency-Shift-Keying (FSK) | ▪ 2-FSK (USFOOK) by Intel Corp., CA 95054-1537 [6]<br>▪ RS-FSK—by National Taiwan University [6]<br>▪ M-FSK by Carnegie Mellon University [10]<br>▪ Compatible M-FSK by Kookmin University [6] | ▪ 2-FSK (UFSOOK) by Intel Corp. [6] |
| Variable Pulse Position Modulation (VPPM) | ▪ PWM code by Panasonic [6]<br>▪ Offset-VPPM by SNUST [6] | ▪ VPPM by Intel Corp. [6] |
| Phase-Shift-Keying (PSK) | ▪ Hybrid FSK/PSK by Kookmin University [6] | ▪ Spatial 2-PSK by Kookmin University [6] |

Abundant research has carried out in the last few years on the applications of OWC technology such as indoor localization [14,15] vehicular communication [16,17], vehicle positioning [18,19], resource allocation [20], *etc.* Other techniques based on screen and camera communication for digital signage and other applications have also been proposed, which have not included the idea of OWC technology [21–23]. Considering the frame rate variation along with the simplicity of encoding and decoding procedures we have designed and implemented a novel compatible modulation and coding scheme for image sensor communication which is an extension of our previous work [24].

For the last concern in the modulation and coding scheme, Table 1 gives a survey of related modulation schemes from the IEEE TG7r1 proposals and outside of TG7r1 as well. None of them mentioned how to merge data from incomplete parts that belong to a payload into a complete one. The data fusion technique is meaningful for communication, especially for a rolling shutter camera-based ISC, not only for situations in which the camera frame rate is low and varying but also for a situation in which the distance is too far, causing a small size of the LED coverage on an image and hence a small amount of data received per image. This paper presents the concept of data fusion from a couple of images by using an asynchronous bit which also effectively solves the frame rate variation problem (*i.e.*, very wide variations). The leading contributions of our novel ideas are presented briefly in the following section.

**Table 2.** Highlights of our contributions.

| Contribution | Necessity | Effectiveness |
|---|---|---|
| **Proposed data frame structure #1 consists of**<br>▪ A short preamble<br>▪ Two asynchronous bits (two Ab bits, one before and one after the payload) | ▪ A short preamble is for OOK modulation scheme and RLL coding.<br>▪ Frame rate variation: The challenge of frame rate variation is simply resolved by inserting asynchronous bits to a data sub-packet [1].<br>▪ Data fusion: Asynchronous bits are helping a varying frame-rate receiver in asynchronous decoding and more, allowing a couple times better performance of transmission (benefits in distance and sub-packet length). | Overhead [3] includes a short preamble and two asynchronous bits at the beginning and the ending of the sub-packet. |
| **Proposed data frame structure #2 consists of**<br>▪ A short preamble<br>▪ Four asynchronous bits (two bits $Ab_1$ and two bits $Ab_2$) | ▪ All above<br>▪ Missed Frame Detection: By using two bits overhead additionally, a receiver is able to detect the missing of three adjacent packets frames. It means error is detectable completely when the frame rate is no less than a quarter of packet rate [2]. | Two bits overhead additionally. |

[1,2] The definition of a packet, sub-packet, packet rate and the other terms are given in Table 3; [3] The overhead which is considered here is the amount of overhead from the PHY layer.

## 2.2. Our Contributions

Different from the other related works mentioned in the subsection above, this paper addresses several novelties and contributions as Table 2 concludes. The first proposed frame structure which consists of a couple of asynchronous bits which is meaningful in supporting two critical functionalities:

a) Frame rate variation support

- The Ab couple allows any receiver which has any frame rate greater than the transmitting packet rate, to decode data in a simple way by checking the value of the bits. Therefore, the wide range in possible frame rates supports greater compatibility with the majority of commercial cameras available on the market.

b) Data fusion algorithm

Asynchronous bits are not only for helping a varying frame-rate receiver in asynchronous decoding, but also for achieving better (by two-fold) transmission performance by using a data fusion algorithm. The idea of our fusing algorithm is to fuse two incomplete data parts from two images into a complete data packet. Compared to a similar work from PureLiFi [9] in which no fusion is applied, fusing two parts of a packet is meaningful to:

- Allow transmission of a longer data sub-packet (a couple of times longer) over the same distance.
- Or enable receiving data from a further distance (a couple of times further) using the same sub-packet length (see Table 3 below for the definition of a sub-packet).

Noticeably, the cost of overhead for both functionalities is just a couple of bits per data packet. The reduction of overhead is very essential for communications in which the data rate is limited. Additionally, compared to a cross-frame linear-ensure-code approach for resolving frame rate variation in a work of Microsoft [12], an insertion of a couple of bits is much simpler and causes much less overhead.

The second proposed data frame structure is to target detection of any missing packets. Since the discovery of where an error happens is sufficient for some uses in which, after detecting the error, the application tells the user what to do or identifies the need for subsequent processes, the correction of missed packets is not of interest in this proposed scheme. Consequently, the simplicity and effectiveness of an error detection code become critical.

Comparing to the cross-frame linear-ensure-code approach for resolving frame rate variation proposed in [12] and the erasure code mentioned in Hiraku *et al.*'s paper [25], our method of inserting a couple of additional bits is simpler. Moreover, the effectiveness (*i.e.*, less overhead and more detectable errors) of our error detection in the presence of wide frame rate variations is outstanding. The effectiveness of Microsoft's approach is that the error is detectable when the frame rate is no less than half of packet rate. Meanwhile, the effectiveness of the Hiraku *et al.*'s method comes from the fact that errors are detectable when the missing packet ratio is less than 8/18 and the code rate of a low-density parity check (LDPC) is about 1/2. In comparison, in our scheme errors are detectable when the frame rate is no less than a quarter of the packet rate by using just two bits of additional overhead (Ab2) per packet. Consequently, the loss of three adjacent packets becomes completely recoverable. However, the definitions used throughout this paper is summarized in Table 3.

**Table 3.** Definitions used in our temporal encoding schemes.

| Term (Synopsis) | Definition | Description |
|---|---|---|
| Data Sub-packet (DS) | A pack of information clocked out, including a preamble, asynchronous bits, and a payload. A sub-packet is counted from the starting of a preamble symbol to the starting of the next preamble symbol. A data sub-packet is shortly called a sub-packet, and being denoted as DS. | Figure 4 |
| Data packet | The set of data sub-packets which includes a sub-packet and its multiple-times-repetition. | Figure 4 |
| Packet rate | The number of different data packets across the transmission medium per time. It shows how fast the data packet is clocked out (e.g., 10 packets/s). A data packet consists of multiple DS. All those DS are same and each of them brings a similar payload. | Figures 4 and 5 |
| DS rate | The frequency at which the DS is clocked out to the transmission medium. | Figure 4 |
| Start Frame (SF) | The pre-defined symbol to be inserted at the starting of every sub-packet. The SF is a preamble which is same on every sub-packet DS. | Figure 4 |
| DS payload | A DS payload is the amount of data (the body data) of a data sub-packet DS. DS payload is also shorten as the payload. A payload is encoded along with asynchronous bits and a preamble to form a sub-packet. A data packet has multiple sub-packets but all those sub-packet bring the same payload. | Figures 4 and 5 |

| Clock information (of a data packet) | The information represents the state of a packet clocked out. The clock information is transmitted along with a DS payload in every sub-packet to help a receiver in identifying an arrival state of the new packet under the presence of the frame rate variation. | Figure 5 |
|---|---|---|
| Asynchronous bit (Ab) | A form of the clock information in the temporal scheme helping a varying frame rate receiver in asynchronous decoding. In our scheme, a single or a couple of asynchronous bit(s) is used representing the clock information. However, note that the clock information is not only necessarily one single bit (bit 1 or bit 0), but also can be a symbol (a set of bits in which symbol 1 and symbol 0 are orthogonal somehow) to operate at high noise affected. | Figure 5 |
| Over-sampling scheme | The communication scheme in which the frame rate of a rolling shutter camera is greater than the packet rate of the transmission. | Figures 6 and 7 |
| Data fusion | A process in asynchronous decoding to group two or more than two incomplete data parts (forward parts and backward parts) those belong to one complete packet. Two statements of data fusion include:<br>• Inter-frame fusion: To group and fuse data parts from different images<br>Intra-frame fusion: To group and fuse data parts from an image | Figures 7 |
| Asynchronous decoding | The decoding procedure under mismatched frame rates between a transmitter and a receiver due to frame rate variation. The initial step of decoding procedure is to detect the SF symbol. From the location of the SF, two sub-decoding procedures are:<br>• Forward decoding: To decode the forward part of the image from the location of the SF.<br>• Backward decoding: To decode the backward part of the image from the location of the SF. | Figure 8 |
| Under-sampling scheme | The communication scheme in which the frame rate of a camera is permitted dropping to less than the packet rate of transmission. Any missed packet due to the frame rate dropping will be detected for a future process. | Figures 9 and 11 |

## 3. System Considerations and Necessity of Our Work

### 3.1. System Architecture and Analysis

Highlighted by the latest version of the TCD, the frame rate variation is considered individually for ISC. In addition, the meeting to call for proposals in January 2015 attracted a lot of interest from several companies and universities [5]. Moreover, the prototype for a CMOS rolling shutter camera has become a technology of interest because of the inexpensive hardware and ubiquitousness of CMOS cameras compared to the global shutter camera series. Consequently, this has created a demand for a compatible modulation scheme for frame rate variations. We have discussed the overall system architecture for a compatible modulation scheme as well as the operation of rolling shutter cameras in this section. We have analyzed the frame rate variation issue by using experimental results. The proposed architecture of the system uses two main parts, the transmitter PHY, and the receiver PHY, for modulating and demodulating data, respectively, to achieve compatibility.

The system architecture shown in Figure 1 uses three fundamental working principles to work as a compatible modulation scheme in the time domain. Our proposed system consists of the transmitter PHY layer, which is used for operating different types of light emitting devices. This PHY layer is connected to a user interface device for transmitting data. However, we have illustrated a novel idea for using a compatibility support module in the transmitter PHY while it is connected to the data packet modulator. This module can support three functions for ISC, which are: (a) frame rate variation; (b) different shutter speeds and sampling rates; and (c) various resolutions and changeable distances. The second fundamental characteristic of our proposed architecture is the data frame format to make the compatibility support modular. The significance of the data frame format is that it contains the clock information along with the data to the receiver, and the data packet is modulated in the time domain. The frame rate variation can cause an error, which can be resolved by the compatibility support. Error detection is another important feature of the proposed architecture.

Thus, the receiver PHY layer needs a compatibility support module that can be operated simultaneously for the data packet demodulator and error detection. The clock information is used to detect the error. A broad discussion on the data frame structure and error detection will be presented in the next section. But before getting detail into the system and data frame structure design, the necessity of our work is going to be discussed in the very next subsection right below.

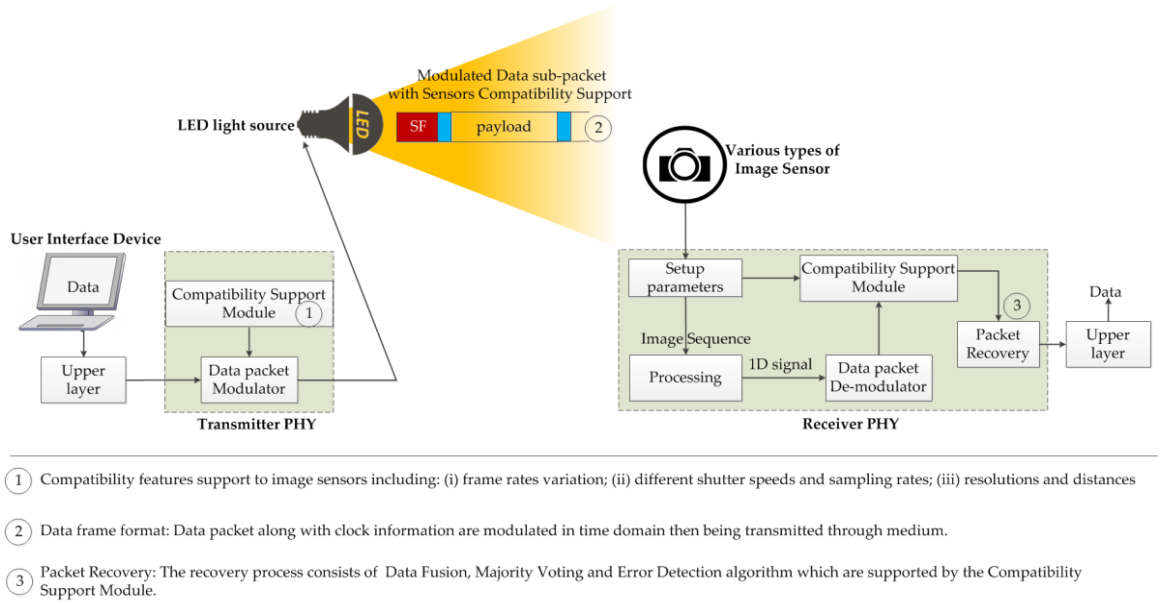

**Figure 1.** System architecture for compatible encoding scheme in time domain.

### 3.2. Necessity of Our Work

#### 3.2.1. Necessity of Data Fusion

The idea of our data fusion algorithm is to fuse two incomplete data parts which belong to a data packet into a complete one. A camera receiver can record data on the capture time of every single image (*i.e.*, the time from the first row to the last row of an image is recorded with a rolling shutter camera, letting us define this the capture time of an image); however, any information on the duration between two adjacent images is lost. Two logics are encouraging us to propose a data fusion given as follows.

The first reason is that a rolling shutter camera may have a different time in capturing a single image from other rolling shutter cameras. Thanks to Tsai *et al.*, a helpful survey on rolling shutter cameras was provided in [26] as a proof, which reached a similar conclusion as us about the variety of rolling shutter parameters. Consequently, between different cameras, there occurs a difference amounting to data recorded per image at full capacity (i.e., at a near distance, an entire image is exposed by LED light). Figure 2a shows an example of the maximum amount of data recorded per image of a camera. Long data packets can be recorded wholly on an image of a long-capture-time camera but remains incomplete on an image of another camera with short-capture-time. The fusion of data from incomplete data parts on images into a complete one is to support plenty of rolling shutter cameras as a key compatibility feature in our proposed system.

The second reason is that on the same camera, the amount of data recorded in an image is inversely proportional to the distance from the camera to the transmitter. Figure 2 clearly shows that the decodable data part of an image depends on the size of the transmitter captured on the image. Consequently, at a further distance where a packet is incompletely captured, a data fusion to merge those data parts into a complete packet is meaningful. From the couple of reasons highlighted, our proposed data fusion algorithm aims to achieve a better performance in communications, allowing the transmission of a longer data frame (a couple of times longer) at the same distance, or allowing the data reception from a further distance (a couple times further) at the same packet length.

Figure 2 shows a test of the operation of rolling shutter to exposure light, a Pulse Width Modulation (PWM) banding operation produces a limit on the maximum amount of data being recovered per image (see Figure 2). When the interval between images is inconsistent, a packet of data is recoverable only if the packet is within an image. The fusion of two discrete parts of a data packet gives us another chance to recover a long data packet from different images.

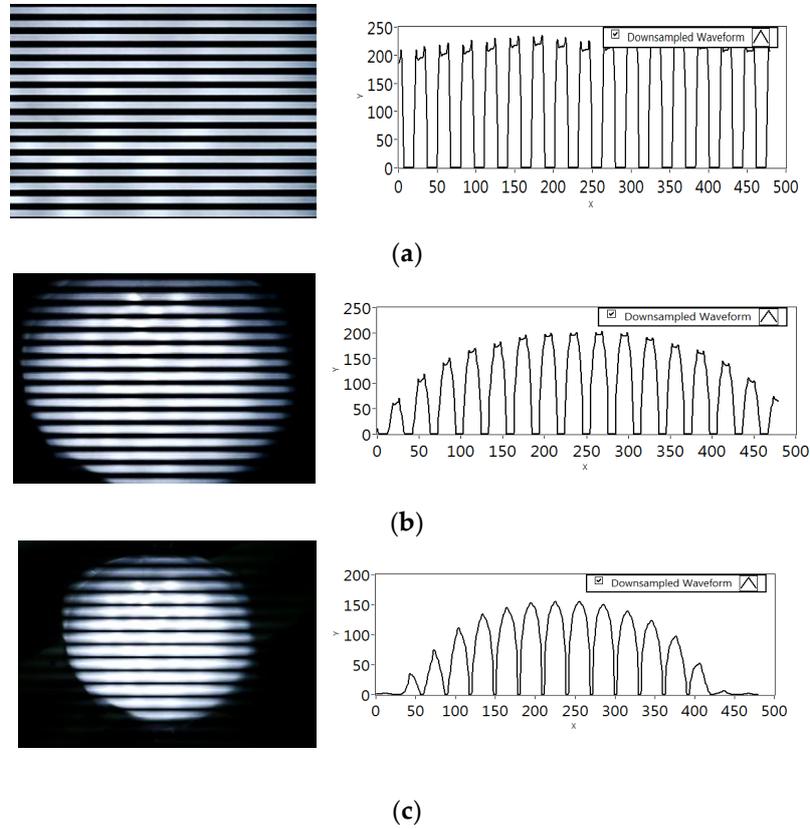

**Figure 2.** A PWM banding experiment. The size of the image of the circle LED (180mm diameter) is decreased meaning that the amount of recoverable data is reduced when the distance is increased from (a) 20cm ; (b) 40cm ; (c) 60cm.

### 3.2.2. Necessity of Frame Rate Variation Support

It had been believed that commercial cameras operate at a fixed frame rate, usually at 30 fps. However, the frame rates of several cameras have been estimated by measuring their inter-frame intervals, and the result of our experiments shown in Figure 3 lead us to conclude that the camera's frame rate always varies. In addition, the experimental result of Figure 3 indicates that the frame rate of the cameras used in the experiments ranges from 20 fps to 35 fps. The varying frame rate value at the time, *t* is considered as a function of time as follow, and this variation can be expressed by formulating Equation (1):

$$R_{frame}(t) = E[R_{frame}] + \delta(t)(\Delta R_{frame})\tag{1}$$

where $R_{frame}(t)$ denotes the value of the varying frame rate at a given moment, *t* and $\delta(t)$ represent the portion of deviation with respect to its previous frame rate at the time, *t*. The value of $\delta(t)$ lies between −1 to 1 (*i.e.*, $-1 < \delta(t) < 1$). The negative value of $\delta(t)$ indicates the decrease in frame rate from its previous value while the positive one stands for the increment in the frame rate. Besides, the term $E[R_{frame}]$ symbolizes the mean value of the frame rates that occur at a period whereas $\Delta R_{frame}$ signifies the maximum deviation of the frame rates.

However, the significant problem is that the frame rate variation is unpredictable. Consequently, operating with an unpredictable variation in frame rate is a challenge. Whenever the LED transmits data during the time between sampling two images, the camera receiver cannot record the data.

Therefore, data can be lost from the packet if we do not have any information about that time period to recover the data. Furthermore, the fixed pulse rate of the LED can cause heterogeneous mismatching in the frame rates when the frame rate varies. To model the frame rate of an image sensor, a consideration of the wide range variation in frame rate benefits the model in being compatible with plenty of cameras available nowadays.

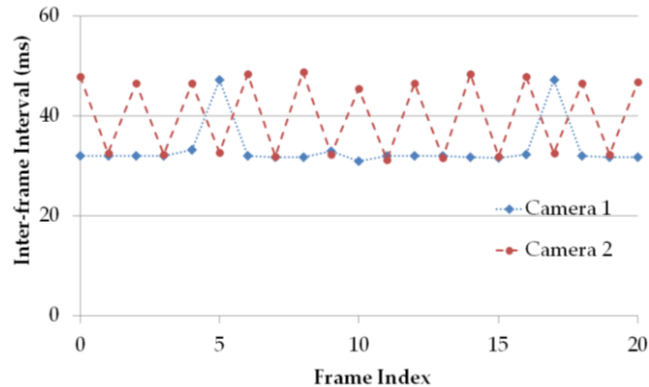

**Figure 3.** Experimental records of cameras' frame rate variation by measuring inter-frame intervals.

### 3.2.3. Necessity of Data Frame Structure

Ever since the final version of the TCD [5] of the TG7r1 came out, the necessity of low-overhead PHY modes has been seriously considered. Each proposer has his preamble symbol (a Start Frame (SF) symbol) to notice the start of a data packet transmission regarding the support for his data reception. A preamble is a frequency delimiter such as proposed in the paper from Intel [7] and Carnegie Mellon University [10] or a set of PWM/PPM symbols as suggested from Panasonic [6]. In comparison, each preamble has its own advantage, being best suitable for the modulation and coding by the proposers. In our rolling shutter ISC system, On-Off-Keying is chosen as the modulation scheme and Run Length Limited (RLL) codes are selected as the coding scheme. Thus, a recommendation of the suitable preamble becomes vital for our data frame structure. The proposed preamble (SF symbol) will be easily distinguishable among RLL data and be short to minimize any additional overhead.

It is readily observable that unidirectional communication is one of the promising services of OWC technology. Furthermore, the cost of the communication technology can be reduced considerably without any uplink requirement in the unidirectional communication, but there has to be an asynchronous scheme on the receiver side. However, the asynchronous method allows a receiver to capture the LED images at any starting time and to recover data by decoding it smartly, even with mismatching frame rates, as mentioned in the previous section. To help the receiver in employing asynchronous decoding, the data frame structure must be specially adapted for ISC technology, and some of the proposed structures have been considered for ISC technology. However, the ISC technology is a revision of the specification of the VLC standard; therefore, the RLL coding schemes for ISC technology can be similar to the VLC technology, which is mainly used in order to maintain an average LED brightness during the communication period. In the VLC specification [4], some well-known line coding schemes including Manchester code, 4B6B code, and 8B10B code have been proposed. These codes are used to maintain brightness constant at 50%, and amplitude modulation (AM) has been used additionally to dim the LED brightness.

The link rate related issues are significant for ISC technology, while the data rate is one the most challenging obstacles that need to be overcome by ISC technology to compete with VLC technology. The data rate efficiency, $\eta$ represents the efficiency rate of actual data rate to link throughput in coding schemes can be computed by Equation (2):

$$\eta = \frac{(Actual\ data\ rate)}{Throughput}.100\%$$

(2)

The data rate efficiencies of the Manchester, 4B6B, and 8B10B codes are compared in Table 4.

**Table 4.** RLL Codes and Data Rate Efficiency.

| Topics | Manchester | 4B6B | 8B10B |
|---|---|---|---|
| Data Rate Efficiency, $\eta$ | 50% | 67% | 80% |
| Shortest length of the preamble required * | 8B | 10B | 18B |

\* The requirement on the preamble length is to ensure the preamble is distinguishable among RLL data. Here, B denotes an OOK state. B can be "1" or "0", which represents LED is on or off respectively.

The results show that the Manchester code is one of the simplest codes, but it has the smallest data rate efficiency. On the other hand, the 8B10B is the most complex code and the most efficient regarding data rate. However, being different from VLC technology, the ISC technology operates within a significantly lower range of frequency due to the limitation of the shutter speed of the camera. Most of the available smartphone cameras have an 8 kHz limit on the shutter speed, thus narrowing the bandwidth for ISC at the same time. The selection of modulation frequency needs to be flicker-free to the human eye while using the limited bandwidth. This restriction further necessitates proposing a new frame structure in order to mitigate LED flickering effectively while still being suitable for the rolling shutter operation of the camera.

## 4. Proposed Schemes

Our proposed schemes can be divided into main categories depending on the frame rate of the camera (receiver), which can be greater or less than the packet rate of the LED (transmitter). One of the schemes is called the oversampling scheme, in which the frame rate of the camera is many times greater than the packet rate of the transmitter, while the other one is known as the undersampling scheme, in which the packet rate is greater than the frame rate of the camera, according to our proposed system. Packet rate is defined as the number of different packets which bring different payloads across the transmission medium per time period. It indicates how fast the data packet is clocked out (e.g., 10 packets/s). However, the definition of a payload of a packet depends on the modulation scheme. According to our proposed plan, the payload is the amount of data within a sub-packet. Every packet can contain multiple sub-packets. However, all sub-packets have the same data payload (see Figure 4). We consider the repetition of data in a packet only once for throughput calculation. With a view to enhancing data rate while keeping the brightness of LED constant, the Manchester, 4B6B, or 8B10B coding are chosen as RLL codes to implement our proposed scheme. Also, the frequency in the range of suitable frequency is used to drive the LED.

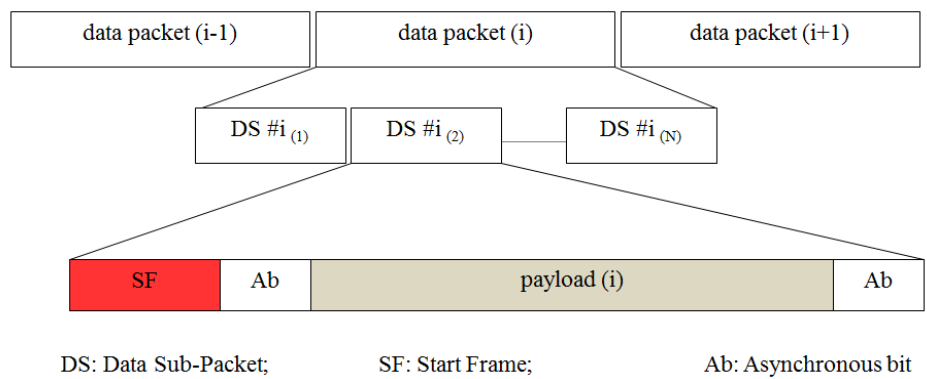

DS: Data Sub-Packet;                SF: Start Frame;                Ab: Asynchronous bit

**Figure 4.** Proposed data frame structure #1 for ISC between rolling shutter camera and LED. A data packet consists of multiple sub-packets which bring the same payload.

*4.1. Oversampling Scheme*

### 4.1.1. Problems and Statements

When the frame rate of a rolling shutter camera becomes several times greater (at least double) than the packet rate of the transmitter, every data packet (a packet has many sub-packets, all sub-packets are same, and each of them carries the same DS payload) needs to be sampled at least twice (*i.e.*, two images). The frame rate variation of the existing rolling shutter camera can cause a problem in the sampling of a DS payload. The compatibility support on modulation and coding can make the procedure capable of decoding without any error. Keeping these issues in mind, we have proposed a novel data frame structure that contains clock information along with a data packet in the form of an asynchronous bit in order to assist the receiver in mitigating the effect of frame rate variation and fusing images data under mismatched frame rates. We have also proposed a majority voting scheme to cancel any errors after grouping different images with the help of the clock information that is discussed later.

### 4.1.2. Data Frame Structure

We have proposed a new data frame structure to support the compatibility of frame rate variation for the oversampling scheme. Our proposed data frame structure consists of several packet frames, which are shown in Figure 4. Every packet contains several repeated data sub-frames (DS), and the DS includes a payload with the start frame (SF) bit and an asynchronous bit (Ab). The asynchronous bit represents the clock information of a data packet, which indicates the state of a new payload is clocked out. The clock information is transmitted along with a DS payload to help a receiver identify the arrival state of a new payload with a variable frame rate. The procedure of combining the asynchronous bit with the payload is illustrated in Figure 5 to form a sub-packet DS.

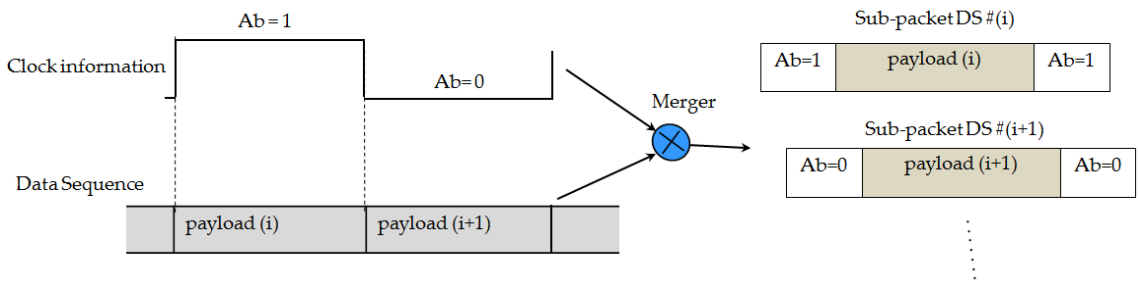

**Figure 5.** Merging procedure of an asynchronous bit and a payload into a sub-packet on time domain.

According to our proposed scheme, every data packet contains a DS in which the DS will be repeatedly transmitted *N* times so that the data will not be lost if it is not captured in the time between images. The proposed structure of the DS with the asynchronous bits included (before and after the payload) helps the camera connect and recover data from different image frames. According to our proposed scheme, each DS frame has two similar asynchronous bits, which is equal to "1" if the number of the packet frame *i* is odd and "0" if *i* is even. In order to avoid missing any of the frames when the frame rate of the camera changes, the value of *N* must satisfy the condition shown in Equation (3):

$$N \geq \frac{\{T_{cam}(t)\}_{\max}}{DS\_length}$$ (3)

where $T_{cam}(t)$ represents the inter-frame interval of a camera during the reception of data while *DS_length* indicates the time duration of the transmission of a DS in a packet. Besides, the function $\{T_{cam}(t)\}_{max}$ calculates the longest inter-frame interval among various inter-frame intervals. The value of *N* needs to be greater than or equal to the ratio of $\{T_{cam}(t)\}_{max}$ to *DS_length* and it needs to be an integer.

### 4.1.3. Statements of Oversampling in Decoding

The decoding procedure depends on the exposure time of the rolling shutter camera (*i.e.*, shutter speed) and interval of the DS in a packet. It is shown in Figure 6 that there exist intervals among the DS in a data packet, and the DS are repeatedly used in the same data packet. However, the DS is encoded with an SF, which represents the start of a DS and the end of an interval between two consecutive DS. There are two cases in the decoding of the oversampled data that are explained below.

- Case 1: Rolling exposure time >> (DS interval)

The oversampled data can be decoded if the rolling exposure time becomes more than twice the value of the DS interval. Consequently, the oversampling happens within every rolling image to recover data in a sub-packet. However, this situation arises when the interval of a sub-packet becomes short compared to the rolling exposure time. Figure 6 shows the oversampled data sub-packet DS in a rolling image with three SF.

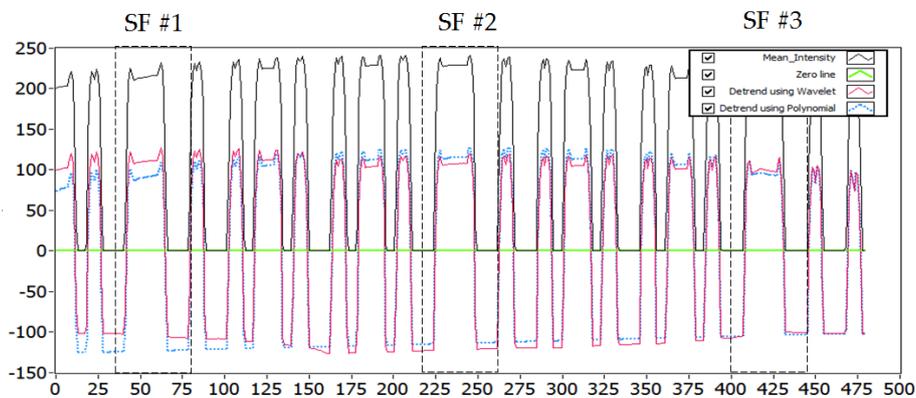

**Figure 6.** An experimental result of oversampled packet in within a rolling image. SF #1, SF #2, and SF #3 are Start Frame of three data sub-packets (three DS).

- Case 2: Rolling exposure time ~ (DS interval)

When the exposure time of a rolling shutter camera becomes approximately equivalent to the time interval between two DS, the rolling image should be able to recover the full length of a sub-packet. This happens when the duration of a DS is so long that it approaches the rolling exposure time. The exposure time of the rolling shutter camera must be greater than (or at least no less than) the time interval of DS because a short exposure time can cause it to miss some part of a DS. However, we have proposed two types of fusion algorithm for recovering data symbols at different sampling times, and they are called inter-frame and intra-frame data fusion.

1. *Inter-frame data fusion*: Fusing two sub-parts of a payload at two different images into a complete payload.

2. *Intra-frame data fusion*: Recovering a complete payload from an image.

A fusion algorithm has been proposed in order to find a lost part from the same image or another image to fuse them together. The data fusion algorithm has been illustrated in Figure 7. The DS is repeatedly transmitted during the transmission of a packet according to our proposed scheme. The idea of the Ab has made it possible to combine the data from different DSs to form a complete payload. In the inter-frame fusing method, the separated data parts from an image are given a different Ab like '1' and '0'. The missing part for each of those incomplete parts (*i.e.*, the forward part and the backward part which are decoded from an image) will be found on the nearest image, and then can be recovered completely by fusing data with the same Ab as the length of DS does not vary. Similarly, this method can be applied to extract data from an image when frame rate varies. The data parts (forward parts and backward parts) that have the same Ab can be fused together. In contrast, the intra-frame fusion is easier when employing the fusion completely in within an image.

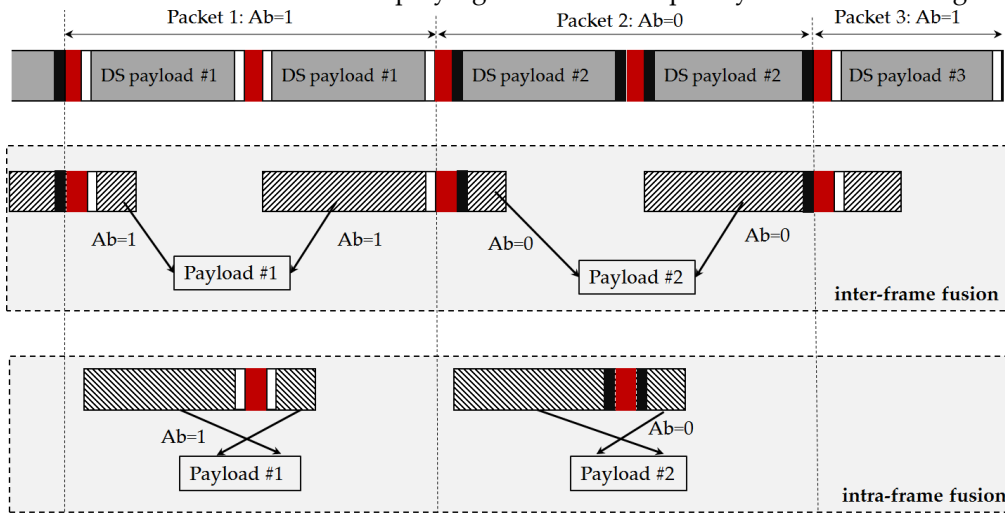

**Figure 7.** Data Fusion Algorithm. *Inter-frame fusion*: to group and fuse data parts from different images into a DS payload; *Intra-frame fusion*: to group and fuse data parts from an image into a DS payload.

### 4.1.4. Proposed Decoding Procedure and Data Frame Recovery

The decoding and the data recovery procedure of our proposed scheme are shown in Figure 8. The decoding method starts by downsampling of an image. After receiving the image, each image frame is processed to achieve down-sampling. The downsampled image is de-trended in order to get the output of a processed signal for SF detection. The next step begins with the demodulation of the image. The modulated process can be divided into two parallel processes, forward and backward decoding, which can be deployed from the position of the detected SF. Two Ab close to the SF are used in DS according to our proposed scheme. These two Ab are used in identifying the data to recover as proposed in Figure 7. Furthermore, the Ab are also utilized for the majority voting scheme to correct any mismatched data errors. The voting is a kind of error correction that is very advantageous in the presence of mismatched frame rates. However, the data can be recovered quickly if and only if every image is captured in a frame. The condition for retrieving the data correctly is given in Equation (4):

$$1 \leq N_{frame} = \frac{t_{cap}}{DS\_length} < N \qquad (4)$$

where the number of images of data captured by each frame is denoted as $N_{frame}$ and $N_{frame}$ is a ratio of $t_{cap}$ and $DS\_length$ while $t_{cap}$ represents the capturing time, which is the time for exposure to one image frame and $DS\_length$ is the time duration of the transmission of a DS in a packet. Besides, $N$ stands for the number of times DS is repeated in a data packet.

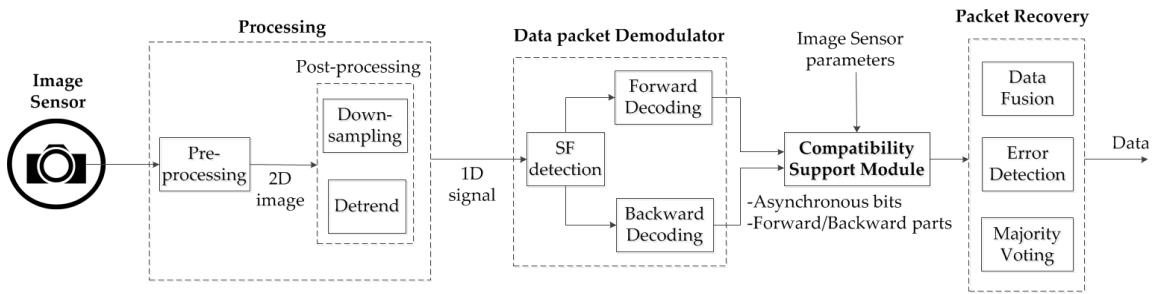

**Figure 8.** Asynchronous decoding procedure using asynchronous bits.

However, the proposed data frame recovery algorithm using asynchronous bits helps the receiver in decoding the whole image frame for recovering data, thus utilizing the data rate to the fullest in unidirectional communication when $N_{frame}$ is considered to be 1. This means 100% the size of the LED is decoded into information without any repetition. In order to get the maximum performance, the whole part of the image has to be decoded and recovered efficiently to obtain one frame of data.

*4.2. Undersampling Scheme with Error Detection*

### 4.2.1. Problems and Statements

The condition for the oversampling scheme is that the frame rate of the camera cannot be less than the packet rate of the transmitter in any situation. For example, if the transmitter operates at 10 packets per second, the frame rate of the camera may vary during the period, but the minimum value of the frame rate of the camera must be more than 10 fps. However, if the frame rate drops to below the packet rate of the transmission while receiving data, a payload might be lost. This phenomenon of the problem in detecting payloads in the undersampling scheme has been explained in Figure 9b. When the frame rate is equal to or greater than the packet rate, the camera can capture at least one payload with the same Ab, and this is illustrated in Figure 9a. Although some part of a DS payload cannot be achieved due to the frame rate variation, the Ab can help us detect the payload by fusing the data of two incomplete payloads with the same Ab. The problem arises when the frame rate becomes less than the packet rate. Some frame can miss capturing the payload as it gets clocked out even multiple times repeated. The missing payload (due to the fact the whole packet is lost) can create an error in grouping two adjacent DSs as well, which can result in an erroneous voting for these two payloads. Figure 9b illustrates the problems that persist in case of the detection of DS in spite of the use of Ab in every sample.

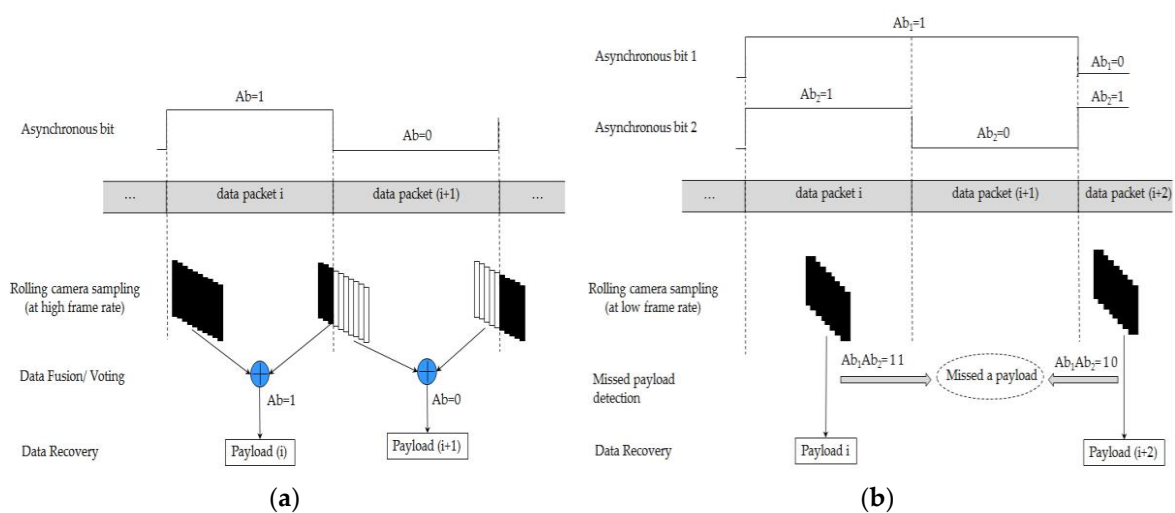

**Figure 9.** Decoding procedures in proposed schemes: (**a**) Data recovery method under frame rate variation without undersampling problem; (**b**) Error in grouping images during undersampling problem.

The Ab used is either 1 or 0 for consecutive DSs. When the frame rate is equal to or greater than the packet rate, the camera can capture at least one payload with the same Ab. The problem arises when the frame rate becomes less than the packet rate. The payload ($i + 1$) with $Ab_2 = 0$ is lost *i.e.*, the camera cannot capture any part of the packet ($i + 1$). Consequently, an error arises while grouping the images to recover the data. It is obvious from Figure 9 that it is not feasible to recover the missing data with the proposed data frame structure of the oversampling scheme. Thus, a new data frame structure is needed for the undersampling scheme which can detect the missing data for the grouping of related images in a proper way. The second proposed data frame is discussed below.

We thank Rick Roberts from Intel for his question and suggestion in the January 2016 Proposal presentation of the TG7r1 about the necessary of error detection mode in case the frame rate of a camera drops to less than the packet rate as an unexpected event. However, the proposed undersampling scheme, as well as the proposed oversampling scheme, have a great impact on our unidirectional mode for ISC technology as the frame rate can go up or down from the desired frame rate at any instant.

### 4.2.2. Data Frame Structure

In order to detect a missing payload, a frame structure similar to the previous one has been proposed and is depicted in Figure 10. The main difference between the frame structures is the use of the Ab. According to our proposed scheme, two Ab have been used instead of one for the undersampling scheme. The data packet contains the similar multiple DS while a different structure of the DS has been proposed. Here, there are two Ab (*i.e.*, $Ab_1$ and $Ab_2$) after the SF and before the data and two Ab after the data in the DS. The first asynchronous bit ($Ab_1$) indicates the packet clock out time, which is similar to the previous packet clock out. The second asynchronous bit ($Ab_2$) indicates a clock out time greater than twice the $Ab_1$. Consequently, the $Ab_2$ can cover both DS intervals for Ab = 1 and Ab = 0. The missing DS payload detection procedure has been explained in the decoding procedure below.

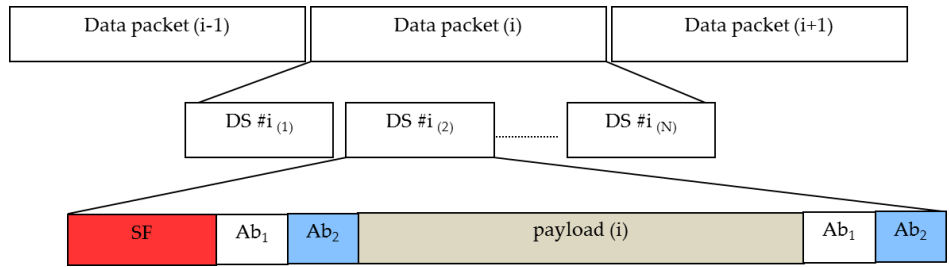

**Figure 10.** Proposed data frame structure #2—a solution for high packet rate transmission or frame drop error correction.

### 4.2.3. Detection of Missed Payload

The missing payload detection is one of the significant parts of decoding procedure as the decoding is complete if and only if the missing payload is detected. The detection of the missing payload is described in Figure 11.

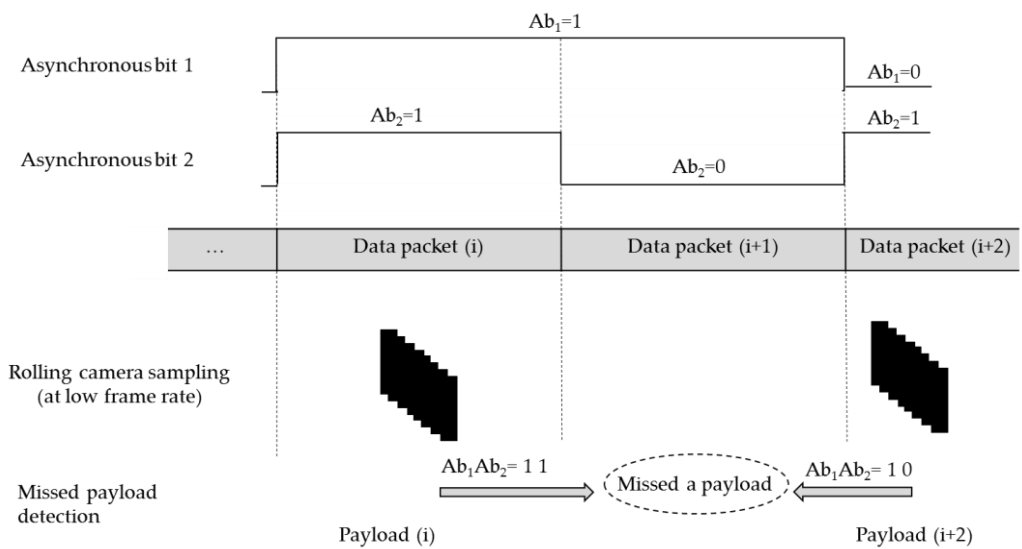

**Figure 11.** Detecting a missed payload from a data packet by using two asynchronous bits.

The use of two Ab (Ab$_1$ and Ab$_2$) can make the detection possible. The first Ab (Ab$_1$) clocks out in half the time of the second Ab (Ab$_2$). The data frame retrieved from the payload $i$ represents the Ab as 11 according to Figure 11. The next data frame from the sampled image indicates that the Ab is 10, but the actual transmitted data frame carries the Ab 01. Thus, the payload ($i$ + 1) is missed, and the corresponding payload can be detected by comparing the asynchronous bits of the two adjacent DSs. However, two Ab generate four different states. Consequently, three missing payloads of the transmitted data packet can be detected by a single receiver simultaneously, which can result in the escalation of packet rate. Once the error is successfully detected, the error correction procedure becomes simple.

## 5. Numerical Results and Analysis

### 5.1. Data Frame and Bit Rate Analysis

The parameters that are considered and used in the experiments of our proposed scheme are listed in Table 5. The features supported by our proposed scheme are recorded as well. The first proposed data frame structure has used only one Ab while the considered varying frame rate is more than or equal to 20 frames per second (fps). According to our proposed oversampling scheme, the frame rate of the receiver must be greater than or equal to the packet rate of the transmitter. The maximum transmitted packet rate used in our experiment to satisfy the condition of the oversampling scheme is 20 packets per second. The output of the experimental results has detected no error due to the variation of frame rate. The use of Ab and data fusing algorithm can mitigate the problem of varying frame rate. Furthermore, another proposed frame structure uses two Ab that can ensure the detection of missing payload in the case of undersampling (*i.e.*, the frame rate is lower than the sampling rate). The assumed frame rate is more than or equal to 10 fps in this frame structure. However, using two Ab can only detect three missing payloads simultaneously. The error is corrected after identifying the missing payloads by outer coding, which is beyond the scope of discussion.

**Table 5.** Experimental Parameters for the Proposed Frame Structure.

| Types of Frame Structure | Receiver's Frame Rate (Assumption) | Transmitter's Packet Rate (Maximum) | Features Support |
|---|---|---|---|
| Frame Structure #1—One Ab | ▪ No less than 20 fps | 20 (packets/s) | ▪ No error detection<br>▪ Error is mitigated by oversampling together with a majority voting<br>▪ Data fusion |
| Frame Structure #2—Couple of Ab | ▪ No less than 5 fps | 20 (packets/s) | ▪ Error detection: able of detecting all missed payloads when the frame rate is no less than a quarter of packet rate.<br>▪ Data fusion |

One of the chief purposes of any communication technology is to enhance the data rate. The main goal of our proposed scheme is to make communication possible with the varying frame rate of a camera receiver. Nevertheless, increasing the data rate is also one of the purposes of our proposed scheme. Therefore, the evaluation of data rate plays an important role. However, the bit rate can be calculated as shown in Equation (5):

$$R_{bits} \leq \eta \left[ \frac{L}{N_{frame}} - OH \right] \{R_{frame}\}_{\min} \qquad (5)$$

where $R_{bits}$ stands for the bit rate and $\eta$ is the data rate efficiency achieved using RLL coding scheme. The term $\{R_{frame}\}_{min}$ denotes the minimum value of video frame rate, and $N_{frame}$ is the number of DS frames captured per image as well as *L* represents the number of states of LED that can be recoded per image. Besides, *OH* indicates the overhead bits (including SF and Ab per image). Our implementation work of the proposed scheme satisfies the condition of $N_{frame} \geq 1$, which indicates that the exposure time of the camera is greater than the length of DS.

The throughput of any communication system can be expressed using the packet rate of transmission instead of the parameters of the varying frame rate as shown in Equation (6):

$$R_{bits} = \eta \left[ L_{symbol} - OH \right] \times R_{packet} \tag{6}$$

where $R_{packet}$ denotes the packet rate of transmission and $L_{symbol}$ represents the data payload per image. It is obvious from Equation (6) that repeated code of any payload during the transmission of data brings no benefit to the data throughput; hence a packet is counted as one payload in terms of data amount.

Practically, the least of the varying frame rates has been found as $\{R_{frame}\}_{min} = 20$ fps in our various experiments. Moreover, the value of $N_{frame}$ is taken as 1 in the experiments that were some of our best trials. The value $N_{frame} = 1$ indicates that the frame length equals the exposure time of the camera, and one image can recover one DS frame accurately. By substituting the values of $\{R_{frame}\}_{min}$ and $N_{frame}$ in Equation (5), we can get the form of a equation for calculating the data rate which is expressed in Equation (7):

$$R_{bits} \leq \eta \left[ L - OH \right] \{ R_{frame} \}_{min} = 20\eta \left[ L - OH \right] \tag{7}$$

The parameters of $\eta$ and *OH* for each coding scheme are given in Table 6 according to our proposed scheme. The value of *L* has been calculated depending on the modulation frequency that has been estimated for the respective data transmission experiment. The formula for calculating the value of *L*, which is an empirical formula, is given by Equation (8):

$$L = \lfloor 0,0311 \cdot f \rfloor \tag{8}$$

where *f* denotes the frequency used in the modulation of the transmitted data, and $\lfloor x \rfloor$ symbolizes the minimum integer that is larger than *x*.

The relation between the data rate and modulation frequency can be derived by calculating the value of the parameters of Equations (7) and (8). Moreover, it is observed from Equations (7) and (8) that the data rate is proportional to the modulation rate used to drive the LED for data transmission. However, the modulation rate must be within the limited frequency range due to the physical limitation of camera shutter speed.

**Table 6.** Coding schemes and overhead.

| RLL Code | Overhead (OH) | |
| --- | --- | --- |
| | *Proposed Preamble (SF)* | *Asynchronous Bit* |
| Manchester | 011100 | |
| 4B6B | 0011111000 | 2 bits/4 bits per packet |
| 8B10B | 0000111111111100000 | |

*5.2. Evaluation of Data Frame Structure #1*

5.2.1. Grouping Using an Asynchronous Bit

Indentifying the values of asynchronous bits is the indispensable initialstep before starting asynchronous decoding. This is because the grouping of incomplete parts which belong to a payload properly is crucial. Our experiments show the presence of an image captured on the transition of two packets. The decoded backward data part belongs to a payload; however the decoded forward data part belongs to the next payload. Fortunately, an asynchronous bit along with the backward part as well as with the forward part completely resolves the grouping matter. Figure 12 and Figure 13 show additional experiments for illustation while Figure 14 shows the practical result when comparing the asynchronous bit of the backward part and the bit of the forward part. As can be seen, those two asynchronous bits are different when an image is captured at the time between two data packets.

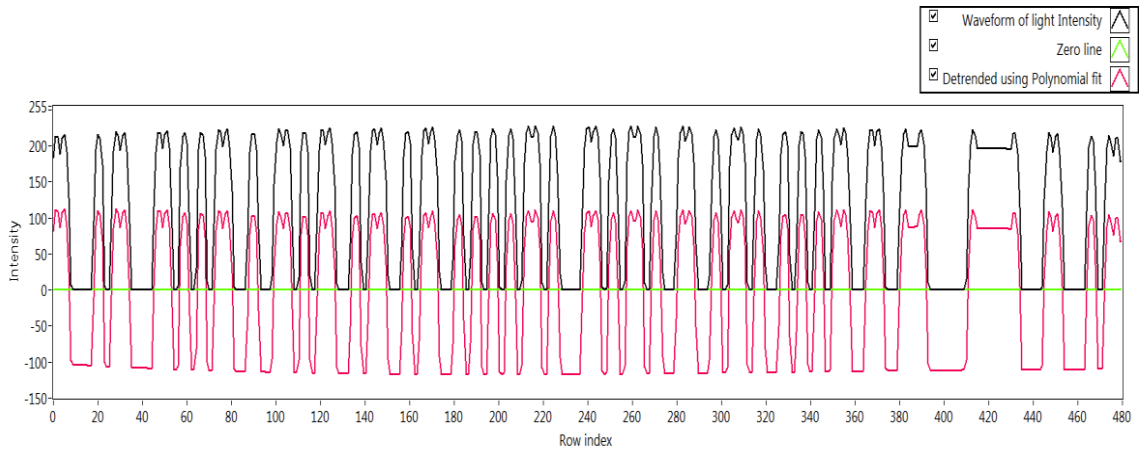

**Figure 12.** An experimental result in asynchronous decoding under mismatched frame rates.

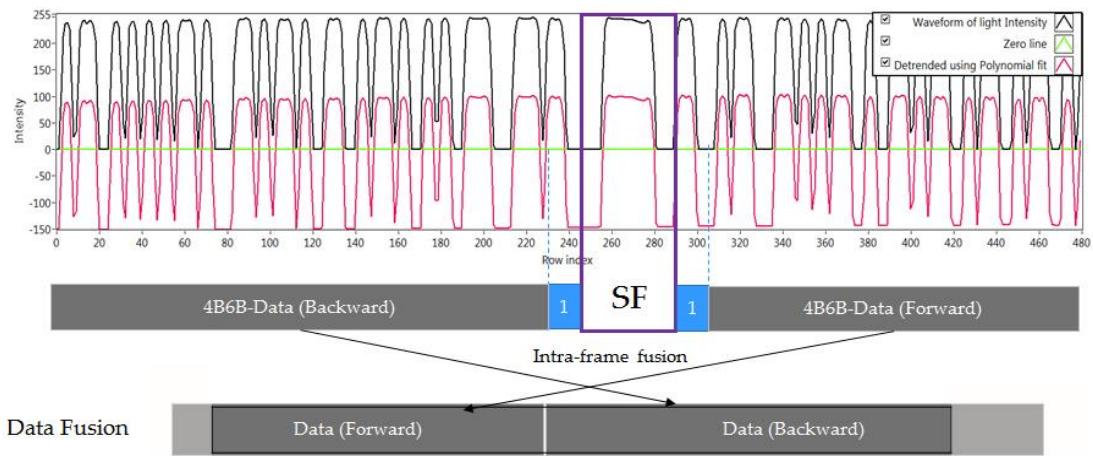

**Figure 13.** Another experimental result with data fusion applied. Here, intra-frame fusion is applied (in within an image) because the asynchronous bits are equal.

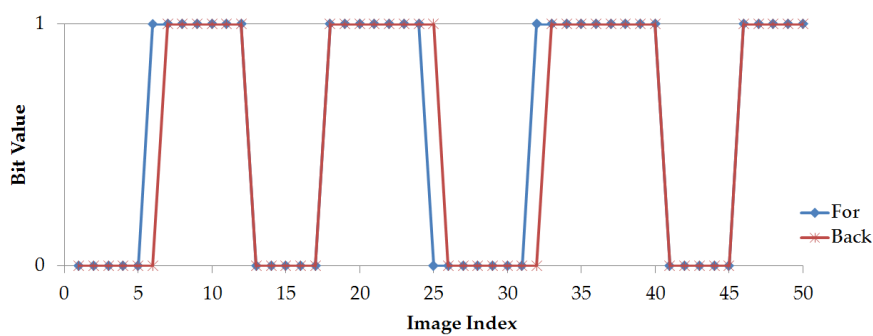

**Figure 14.** The experimental relation between the asynchronous bit of the forward part and the bit of the backward part. The packet rate at 5 packet/s is used for an illustrative graph.

### 5.2.2. Majority Voting Using Asynchronous Bits

Identifying the values of asynchronous bits is the vital initial step before starting asynchronous decoding. Because the grouping of incomplete parts which belong to a payload properly is crucial. Our experiments show the presence of an image captured on the transition of two packets. The decoded backward data part belongs to a payload; however the decoded forward data part belongs to the next payload. Fortunately, an asynchronous bit along with the backward part as well as with the forward part ultimately resolves the grouping matter. Figure 12 and Figure 13 show additional experiments for illustration while Figure 14 shows the practical result when comparing the asynchronous bit of the backward part and the bit of the forward part. As can be seen, those two asynchronous bits are different when an image is captured at the time between two data packets.

**Table 7.** An experiment of grouping incomplete data parts and voting (packet rate = 5 packets/s).

| Ab | 1 | 1 | 1 | 1 | 1 | 1 | 1 | 1 | 0 | 0 | 0 | 0 | 0 | 0 | 1 | 1 | 1 | 1 | 1 | 1 | 1 | 1 | 0 | 0 | 0 | 0 | 0 | 0 |
|---|---|---|---|---|---|---|---|---|---|---|---|---|---|---|---|---|---|---|---|---|---|---|---|---|---|---|---|---|
| **Forward Data (Char Type)** | 0123 | 0123 | 01234 | 012345 | 0123456 | 01234567 | 012 | 12 | abcde | abcdef | abcdefg | abcdefg | abcdefg | abcde | 0123 | 0123 | 01234 | 012345 | 0123456 | 01234567 | 012 | 12 | abcde | abcdef | abcdefg | abcdefg- | abcdefg | abcde |
| **Grouping** | 7 samples | | | | | | | | 5 samples | | | | | | 8 samples | | | | | | | | 6 samples | | | | | |
| **Voted Data** | 01234567 | | | | | | | | abdcefg | | | | | | 01234567 | | | | | | | | abcdefg | | | | | |

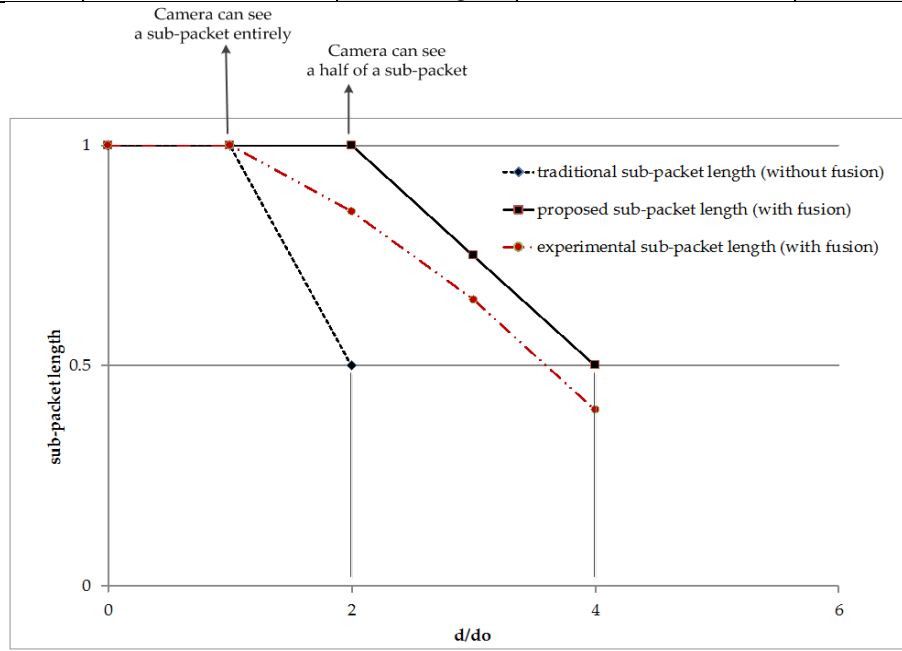

**Figure 15.** Performance comparison between with and without using data fusion approach. *d* denotes the distance between a LED transmitter and a camera receiver. $d_0$ is the maximum distance to get a complete data symbol as expressed in Equation (9). The length of a sub-packet (*DS_length*) is defined in the Table 3.

### 5.2.3. Data Fusion Analysis

Let us assume that there is no data fusion at the beginning of the transmission of data. In this case, the maximum distance of transmission is limited because a camera needs to be close enough to the transmitter to get the full length of the payload (the length of DS) on a rolling image. The maximum distance, $d_0$ is inversely proportional to the sub-packet length and can be expressed as follows:

$$d_0 = \frac{F \times (LED\_size)}{DS\_length} \tag{9}$$

where *F* is the focal length of camera, *LED size* signifies the size of the LED transmitter.

When the two-image data-fusion algorithm is applied, a camera receiver can be further from the transmitter, where it captures half of the payload for merging the data from two images. Therefore, the maximum possible distance for data transmission can be increased a couple of times. At any distance which is less than the maximum allowed distance, the camera can recover the complete payload. At the distance, *d* which is over the limited distance ($d > d_0$), the sub-packet length must be less than the initial sub-packet length in order to be received and recovered. Figure 15 shows the relation between the distance of the transmitter and the receiver and the sub-packet length expressing the expected length and experimental length as well.

As illustrated in Figure 15, the sub-packet length should be reduced by half in order to transmit the packet twice as far. In a comparison between approaches with/without the data fusion algorithm, data fusion is expected to provide twice the performance. It means the same sub-packet length can be transmitted twice the distance or twice of the amount of the data can be received without changing the distance. In practice, however, for a reliable link, we choose the length of a data sub-packet as being less than twice the initial length in order to fuse data efficiently.

### 5.3. Evaluation of Error Detection Frame Structure

#### 5.3.1. Data Rate Analysis with Error Detection

Similar to the oversampling scheme, the data rate is calculated as a function of the packet rate used in transmission as shown in Equation (6). Note that the amount of overhead per data frame is more because of the additionally used couple of asynchronous bits for error detection. In comparison, under the same frame rate variation range of a camera receiver, the maximum capable value of the packet rate in error detection scheme is allowed higher than that of the oversampling scheme. The packet rate is transmitted up to four times of the frame rate of a receiver; hence, more data is clocked out while all missed payloads are being detectable. The tradeoff between the packet rate and overhead shows a significant improvement between the actual data rate of those two schemes.

Equation (10) shows the bit rate, $R_{bits}$ at PHY SAP (Service Access Point) considering the error correction:

$$R_{bits} = \eta \left[ L_{symbol} - OH \right] \times R'_{packet} - OH_{error\_correction} \qquad (10)$$

where $R'_{packet}$ represents the packet rate of transmission of the scheme that is higher than that of the oversampling scheme (equivalent to the packet rate). $OH_{error\_correction}$ symbolizes the amount of overload per sub-frame.

#### 5.3.2. Error Rate Analysis

An evaluation of error probability on frame rate variation is meaningful to reveal whether the proposed error detection scheme is working or not. The error probability also signifies that the data frame structure with couple of asynchronous bits allows the detection of the missed three adjacent packets. Later, the correction of error detection is considered, or simply a notification to the user is shown without the need of error correction. The number of error due to the dropping of frame rate is of interest to reveal our error detection scheme is suitable or not. Therefore, the estimation of error probability in regarding to frame rate variation is meaningful. However, it is explained that our error detection can guarantee that there is no error happens when the frame rate of camera is always greater than a quarter of the packet rate. This occurs because a couple of different asynchronous bits $Ab_1$ and $Ab_2$ can stand for four different states representing the packet is clocked out.

Evidently, a payload might be lost when the entire packet is lost. Hence, the payload can be missed in sampling, and it happens when both of those assumptions occur: (i) the instant sampling rate drops to less than a quarter of the packet rate; and (ii) the next payload is omitted in sampling. However, the probability of a packet is skipped in sampling at $t(i + 1)$ and the probability of non-detected can be calculated as follows:

$$prob(t_{i+1} > 6T \,|\, 0 \le t_i \le T) = prob((t_i + \Delta t) > 6T \,|\, 0 \le t_i \le T) = prob(t_i > (6T - \Delta t) \,|\, 0 \le t_i \le T)$$

$$= \begin{cases} \dfrac{\displaystyle\int_0^T \dfrac{\Delta - (6T - t_i)}{2\Delta} dt_i}{6T}; & if \quad \Delta \ge T - t_i = \dfrac{\displaystyle\int_{T-\Delta}^T \dfrac{\Delta - (6T - t_i)}{2\Delta} dt_i}{6T} = \dfrac{\Delta}{24T} \\[4mm] 0; & if \quad \Delta < T - t_i \end{cases} \tag{11}$$

where $T$ symbolizes the packet length whereas $\Delta$ denotes the amount of the instant sampling interval that is longer than the packet length at the time instance $t_i$.

For expressing the relation between the frame rate variation and the packet rate, an average error rate can be approximately calculated as shown in Equation (12). The rate of improper detection of missed packets can be expressed by DER (Detection Error Rate) as in Equation (12) displays a status of our scheme to ensure that the dropping of frame rate is acceptable or not:

$$DER = \begin{cases} 0; & if \quad R_{frame}(t) \ge \dfrac{R_{packet}}{4} \\[3mm] \dfrac{R_{packet} - \overline{R}_{frame}}{24 R_{packet}^2}; & otherwise \end{cases} \tag{12}$$

where $DER$ is an estimation of the value of error rate in detecting the presence of any missed packet while $R_{packet}$ stands for the packet rate of transmission. Besides, $R_{frame}(t)$ is the varying frame rate at the instant $t$, and $\overline{R}_{frame}$ represents the average value of the frame rate.

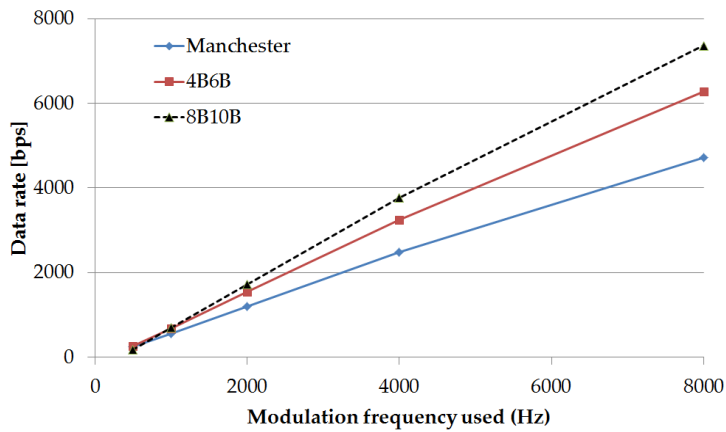

**Figure 16.** An estimation of data rate limit (free-error) in the available frequency range. See Equation (6) for calculation.

## 5.4. Implementation Results

Our experiments have been conducted several times at various modulation frequencies. Different cameras have been tested to verify the variation in the frame rate. The data has been modulated with desired SF and Ab for the asynchronous communication. The asynchronous decoding and the data fusion technique have been depicted graphically in Figure 6 which is the experimental procedure of detecting SF and data processing at the modulation frequency of 1 kHz. In addition, the experimental result of the data rate along with numerous modulation frequencies for different RLL coding (Manchester coding, 4B6B, and 8B10B coding) has been shown in Figure 16. This Figure illustrates that 8B10B coding has the highest data rate at high modulation frequency while the 4B6B has the highest data rate at a very low modulation frequency. Besides, some experimental parameters and achieved results for Manchester coding and 486B coding have been filed in Table 8.The results have been obtained by using regular USB camera while the frame rate variation has been observed as fluctuating between 20 fps and 35 fps. The experiment has been conducted based on an asynchronous scheme. The Table shows that 1.0 kbps data rate has been achieved using Manchester coding at the modulation frequency of 2 kHz with a free error link rate of 1.2 kbps. The data rate of 1.5 kbps can be attained at 2 kHz modulation frequency for 4B6B coding scheme according to our experimental result that has been acquired. The free-error link rate was 1.9 kbps during the experiment for the 4B6B coding at 2 kHz modulation frequency. The experimental results have a clear indication for high data rate communication for rolling shutter camera even though the frame rate of the camera has varied significantly.

**Table 8.** Achieved System Parameters.

| Transmitter Side | | | | |
|---|---|---|---|---|
| **RLL code** | | **Manchester** | | **4B6B** |
| Optical clock rate | | 1 kHz | 2 kHz | 2 kHz |
| Packet rate | | 10 packet/s | | |
| LED type | 15 W | | | |
| **Receiver Side** | | | | |
| Camera type | USB Webcam | | | |
| Camera frame rate | 20 fps to 35 fps | | | |
| **Throughput** | | | | |
| Bit rate limit (free error) | | 0.6 kbps | 1.2 kbps | 1.9 kbps |
| Achieved bit rate throughput (free error) | | 0.3 kbps | 0.5 kbps | 0.6 kbps |

## 6. Conclusions

This paper is an illustration of the compatibility support in the encoding schemes of image sensor communication technology, mostly for the frame rate variation that is present in rolling shutter cameras. The proposed frame structures and decoding systems for the oversampling prototype and undersampling prototype can effectively mitigate the frame rate variation. Besides, high-speed transmission (a few kbps) has been achieved along with the detection of any possible lost payload. However, there is a trade-off between the data rate and distance of transmission. The exposed LED size on an image is tiny at a far distance, hence, providing little data per image. A data fusion which was proposed for merging data from images by using asynchronous bits is a novel idea to recover a complete payload. Moreover, the data fusion technique which combines data from a couple of images into an entire payload enables high communication performance. The same sub-packet length is decodable at two times further distance or the communications data can be double for the same distance. Plenty of useful analysis has been indicated along with the numerical experiments and demonstrated results.

The fusion technique is performed by inserting an asynchronous bit at the beginning and the end of the payload on the sub-packet. In extension, the data fusion method can be applied in merging the data parts from three or more images into a complete payload by inserting an asynchronous bit not only at the beginning and the end of the sub-packet but also in the middle or any place. In this way, an extended data sub-packet can be recovered at a far distance by fusing small parts of the sub-packet from a set of images. An inner code might be generated within a sub-packet for combining plenty of images. In addition, the possible solution for distance transmission has been accomplished by using zoom module in the receiver side. A big size LED can be used on the transmitter side to increase the transmission distance without dropping the data rate too. Suitable RLL coding schemes and the respective SF symbols were proposed for each coding scheme to acquire the high data rate efficiency per data sub-packet while maintaining LED brightness in the communications. Finally, link rate performance has been demonstrated throughout plenty of analysis.

**Acknowledgments:** This research was supported by the Basic Science Research Program through the National Research Foundation of Korea (NRF) funded by the Ministry of Education (No. 2013057922).

**Author Contributions:** Y.M.J supervised whole work; T.N conceived the scheme, T.N and M.A.H designed the experiments; T.N performed the experiments, T.N and M.A.H analyzed the data; M.A.H and T.N wrote the paper; M.A.H and T.N revised the critical correction of the work; Y.M.J contributed to material and tool analysis as well as checked the validity of the experiments.

**Conflicts of Interest:** The authors declare no conflict of interest.